\documentclass[conference]{IEEEtran}
\IEEEoverridecommandlockouts
\usepackage{cite}
\usepackage{amsmath,amssymb,amsfonts}
\usepackage{algorithmic}
\usepackage{graphicx}
\usepackage{textcomp}
\usepackage{xcolor}
\usepackage{stfloats}
\makeatletter

\newcommand{\Rmnum}[1]{\expandafter\@slowromancap\romannumeral #1@}
\makeatother
\usepackage{array}
\usepackage{makecell}
\ifCLASSINFOpdf
\makeatletter
\usepackage{amsmath}
\usepackage{amssymb}
\usepackage{comment}
\def\BibTeX{{\rm B\kern-.05em{\sc i\kern-.025em b}\kern-.08em
    T\kern-.1667em\lower.7ex\hbox{E}\kern-.125emX}}
\begin{document}

\title{SCSGuard: Deep Scam Detection for Ethereum Smart Contracts}

\author{
\IEEEauthorblockN{Huiwen Hu and Yuedong Xu}
\IEEEauthorblockA{School of Information Science and Engineering, Fudan University\\
ydxu@Fudan.edu.cn}
}

\maketitle

\begin{abstract}
Smart contract is the building block of blockchain systems that enables automated peer-to-peer transactions and 
decentralized services. With the increasing popularity of smart contracts, blockchain systems, in particular Ethereum, have been the ``paradise'' of versatile fraud 
activities in which Ponzi, Honeypot and Phishing are the 
prominent ones. Formal verification and symbolic analysis have been employed to combat these destructive scams by 
analyzing the codes and function calls, yet the vulnerability of each \emph{individual} scam should be predefined discreetly. In this work, we present SCSGuard, a novel 
deep learning scam detection framework that harnesses the 
automatically extractable bytecodes of smart contracts 
as their new features. We design a GRU network with attention mechanism to learn from the \emph{N-gram bytecode} patterns, and 
determines whether a smart contract is fraudulent or not. 
Our framework is advantageous over the baseline algorithms in three aspects. Firstly, SCSGuard provides a unified solution to different scam genres, thus relieving the 
need of code analysis skills. Secondly, the inference of SCSGuard is faster than the code analysis by several order of magnitudes. Thirdly, experimental results manifest that SCSGuard achieves high accuracy (0.92$\sim$0.94), precision (0.94$\sim$0.96\%) and recall (0.97$\sim$0.98) for both Ponzi and Honeypot scams under similar settings, and is potentially useful to detect new Phishing smart contracts.
\end{abstract}


\section{Introduction}
Blockchain is an open, distributed and append-only ledger in which all transactions between two parties are recorded verifiably and permanently \cite{Iansiti}. Every block is linked to the previous one via a cryptographic hash, thus forming a chain of blocks. Ethereum \cite{Wood} is one of the largest blockchain system in term of market capitalization, second only to Bitcoin. It is built on top of Bitcoin's principles, but significantly augments the functionality of Bitcoin through ``smart contracts''. 
In fact, the smart contracts are computer programs that are stored and executed via the Ethereum Virtual Machine (EVM), and they can be deployed, invoked, and removed from Ethereum through transactions. However, along with the popularity of smart contracts, Ethereum has been targeted by various severe cybercrimes \cite{Holub} such as the DAO hack in 2016 \cite{Hacking} and the Parity wallet hack in 2017 \cite{Petrov} that 
result in a total loss over 400 million USD. A recent trend on Ethereum smart contract security is the proliferation of less harmful but more latent scams including Ponzi schemes, Honeypots and Phishing scams. Outwardly, they camouflage 
themselves as ordinary smart contracts, while enticing 
benign users to transfer cryptocurrency to the attackers. 
Therefore, finding out the scams from a haystack of smart contracts has become an urgent security issue in Ethereum. 

Symbolic analysis of smart contracts has demonstrated to be valuable in the detection of vulnerabilities. As a pre-deployment bug detector, Oyente \cite{Luu}  uses symbolic execution to capture traces of smart contracts that match the predefined characteristics of vulnerabilities at the bytecode level. However, it is neither sound or complete due to its design flaws that might result in
several false alarms even in simple contracts \cite{kalra1}.
Another tool, Maian \cite{Nkol}, which employs interprocedural symbolic analysis, was able to find many well-known bugs by specifying and reasoning trace properties precisely. Maian classifies vulnerable contracts into three categories - suicidal, prodigal, and greedy. With the bytecode of Ethereum smart contracts, Maian can detect different kinds of vulnerabilities after multiple invocations. However, this code analysis is time consuming, and its accuracy is throttled by the exhaustiveness of the search, i.e. invocation depth, leaving some vulnerabilities uncovered. 
Recently, with the increasing volume of Ethereum transactions, machine learning becomes almost imperative to extract patterns automatically for fraud detection. A tool based on sequence learning algorithms to detect vulnerabilities in smart contracts is discussed in \cite{Tann}. Machine learning method is also used in some specific attack detection tasks such as Ponzi schemes \cite{Chen}.

In this work. we introduce SCSGuard, a framework based on $n$-gram features and attention neural network to detect scams in smart contract using bytecode of contracts which is available for all contracts on the Ethereum platform. Unlike many previous works that have applied static and dynamic analyses to find vulnerabilities in smart contracts, we do not attempt to define any features; instead, we aim to explore methods of machine learning techniques to create accurate models to detect scams in smart contracts at large scale. A significant advantage of SCSGuard is that it requires no specific bug patterns or unique rules defined by human experts. Thus, the labor cost will be reduced considerably. To the best of our knowledge, our proposed method is the first machine learning approach based on $n$-gram features of bytecodes for scam detection in Ethereum smart contracts. SCSGuard applies a sequence learning neural network GRU with an attention layer to catch hidden information in the bytecode of smart contracts.  We evaluate SCSGuard on several well-known scams, and the experimental results show our method performs well in these detection tasks. SCSGuard achieves an accuracy of 0.922 for Ponzi scheme detection and 0.947 for honeypots detection. Other evaluation metrics including precision, recall and F1-score also demonstrate the efficiency of SCSGuard.

The rest of the paper is organized as follows. Section \Rmnum{2} introduces fundamental concepts of our work, including Ethereum platform, scams in smart contracts and $n$ gram method. Section \Rmnum{3} offers the description of our data and features. The implementation details of our proposed model are presented in Section \Rmnum{4} and results of experiments are showed in Section \Rmnum{5}. Finally, we conclude our work in Section \Rmnum{6}.

\section{Background}
\subsection{Ethereum and Smart Contract}
Ethereum allows users to create their own applications within several lines of code, is currently one of the most popular platforms for cryptocurrency based on Ethereum Virtual Machine (EVM), Like Bitcoin, Ethereum also relies on a public, append-only ledger. Transactions in certain periods will be packaged into blocks and then append into the blockchain by miners independently with a consensus algorithm. The fee for the transaction is the user-defined $Gas price$ multiplying the consumed $Gas$ (measure the work a transaction takes to perform). Users and contracts can store money in their addresses or send/receive ether to other users and contracts. Once deployed, no modifications can be made to the contract. 

Smart contracts are programs typically written in a high-level language. The most popular used programming language of smart contract is Solidity. After deploying on the Ethereum blockchain, smart contracts will be compiled into series of EVM executable bytes, which are stored, verified, and executed on the blockchain. EVM bytecode is composed of a series of bytes, each of which is an operation code corresponds with a mnemonic form and a special function. These opcodes represent the operations to be performed. For example, the byte 0x11 corresponds to GT, which means greater-than comparison. The complete form of EVM bytecode and its corresponding opcode is available in the appendix of Ethereum yellow paper \cite{Wood}. 

\subsection{Scams in Smart Contracts}
Recently, smart contracts have seen various adoptions in many domains [10-12], with the rapid development of blockchain technology. Meanwhile, there have also appeared increasingly frauds in the name of digital currency trading because platforms like Ethereum operate in open networks which arbitrary participants can easily join without acquiring permission. The main attacks and threats of smart contracts were discussed in [13]. In addition to smart contracts threats, a series of studies focused on the detection of fraud contracts in the Ethereum platform [14-21]. According to [14], the first empirical study of blockchain financial scams, more than 7 million USD been gathered in only a year. Four categories of scams were defined: Ponzi schemes, mining scams, scam wallets and fraudulent exchanges. Other frauds like honeypots, phishing scams are discussed in [16-21]. These frauds can be quite damaging due to the autonomy and immutability of blockchain system. Thus, it highlights a strong requirement for the detection of such security problems of smart contracts.

We briefly go over some common financial attacks in smart contract, including Ponzi scheme, honeypot and phishing scam in the following.
\subsubsection{Ponzi Scheme}
Ponzi scheme is a kind of classical fraud that appeared almost 100 years ago. It typically uses revenue paid by new investors to generates returns for older investors. However, if there is not enough circulating money, participates, especially those new investors, will lose their money. Ponzi schemes also emerge in blockchain recent years. To deal with the problem that Ponzi schemes usually use many addresses at the same time, in the work of [16], a clustering method to detect Bitcoin Ponzi schemes were proposed. Meanwhile, in the work of [15], a study about Ponzi schemes on Ethereum were conducted, it collected 16 million transactions from July 2015 to May 2017, and a total 17,777 transactions were found to be related to Ponzi schemes.

\subsubsection{Honeypot}
Honeypots smart contracts are contracts that are designed to look like an easy target to attract users. The source code of honeypots is usually publicly available; users will be easily tricked into believing that the contract contains a vulnerability. For example, user can easily be fooled to believe that the contract will return the total amount of its balance after receiving some money. However, the condition to transfer money (msg.value \textgreater \ this.balance ) will never be satisfied, because the balance will increase before the function execution. A symbolic execution to detect smart contract honeypots was used in the work of [17], it also developed a taxonomy of honeypot. In [18], a machine learning approach based on transaction data, source code, and flow of funds was used to detect several specific kinds of honeypots.

\subsubsection{Phishing Scam}
Phishing scam is a form of electronic identity theft using the method of disguising as an honest firm, aiming at obtaining private information such as usernames and passwords from users who visit them [19]. A traditional phishing attack typically begins by sending an electronic letter that seems to be from an authentic organization. This email encourages the victims to click on the provided address, which will further direct them to an illegal website. Recently, phishing scams also emerge in the trading system of Ethereum.

\subsection{N-gram Classification Method}
The $n$-gram model is one of the most common used language models in natural language processing and has been successfully used in many tasks, such as language modeling and speech recognition [21]. N-gram means the N-character slice of a long string, which is a useful substring because all substrings of the file have a fixed length $n$ \cite{Abou-Assaleh}. The Common N-Gram analysis method has been successfully used in many NLP tasks including text clustering, automatic authorship attribution, etc. N-grams can not only capture statistics about substrings of length $n$, but also capture frequencies of longer substrings. This characteristic is noted in NLP, that though 2-grams or 3-grams seem to be too short to be useful, they indeed frequently obtain sound results. 

\begin{figure}
\centering
\includegraphics[scale=0.67]{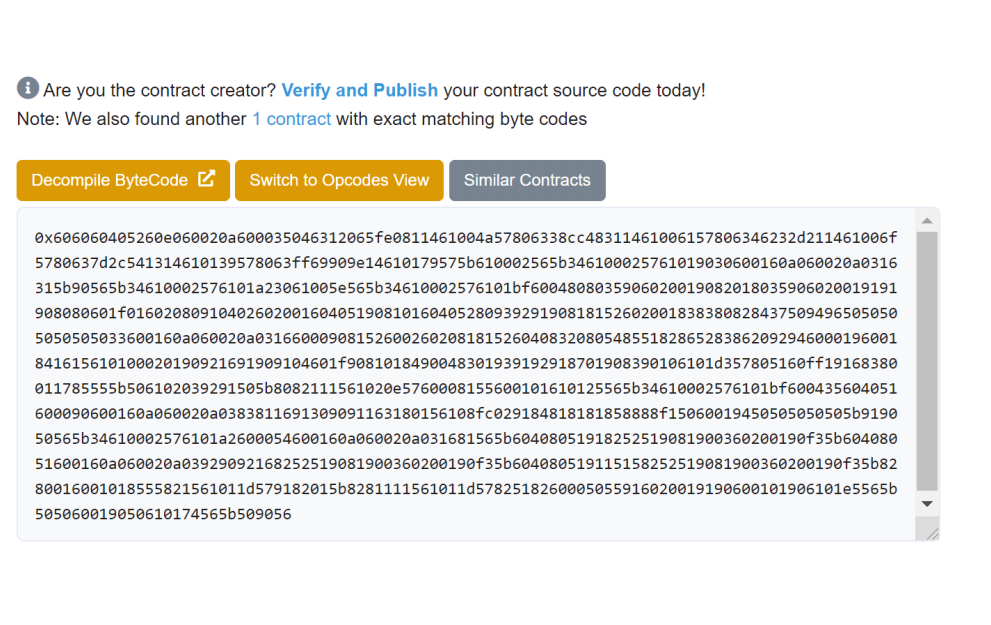}
\caption{The sample bytecode of a smart contract at address 0x01ade83-a7ac7d13ab01f322d68bc2f8fe371ed27.}
\end{figure}

N-grams also perform well in malicious code detection. N-gram could capture specific features according to the coding styles or behaviors of code authors. Besides, many code writers tend to write and compile their code with tools. N-grams could also detect the specific features of certain tools, such as code generators, compilers, etc. Since the captured $n$-gram features are implicit, it would be difficult for virus writers to write viruses without being discovered by $n$-gram analysis deliberately, even though they may know the detection algorithm.

Most of these studies represent the inspected files by extracting byte $n$-gram-pattern features. Encouraged by their promising results, in this study, we also represent $n$-grams of smart contract bytecodes, then use them to capture the inside features of vulnerable smart contracts.

\section{Data and Feature Extraction}
\subsection{Data Source}
The source code of a contract is an essential part to analyze its function and detecting hidden vulnerabilities. However, many smart contracts do not open its source code for public inspection, only less than 1$\%$ of them are open source. Considering this problem, in this work we only rely on bytecodes of smart contracts, which are publicly available for any contract to be implemented by the Ethereum blockchain.

First, we collect addresses of the verified smart contracts reported by previous studies \cite{Bartoletti}\cite{Camino}. With an attempt to evaluate our proposed model, we collected three kinds of scam-related contracts: honeypots, Ponzi schemes and phishing scams. We collect 412 verified honeypots addresses, 183 verified Ponzi schemes addresses, and 7799 verified safe smart contracts addresses in all. Then, we use Etherscan Website http://etherscan.io, which is a block explorer and analytics platform for Ethereum, to automatically collect the bytecode of these verified smart contracts. As is seen in Figure 1, given the address of a smart contract, by using Etherscan, we can get its bytecode information easily. 

\begin{figure}
\centering
\includegraphics[scale=0.7]{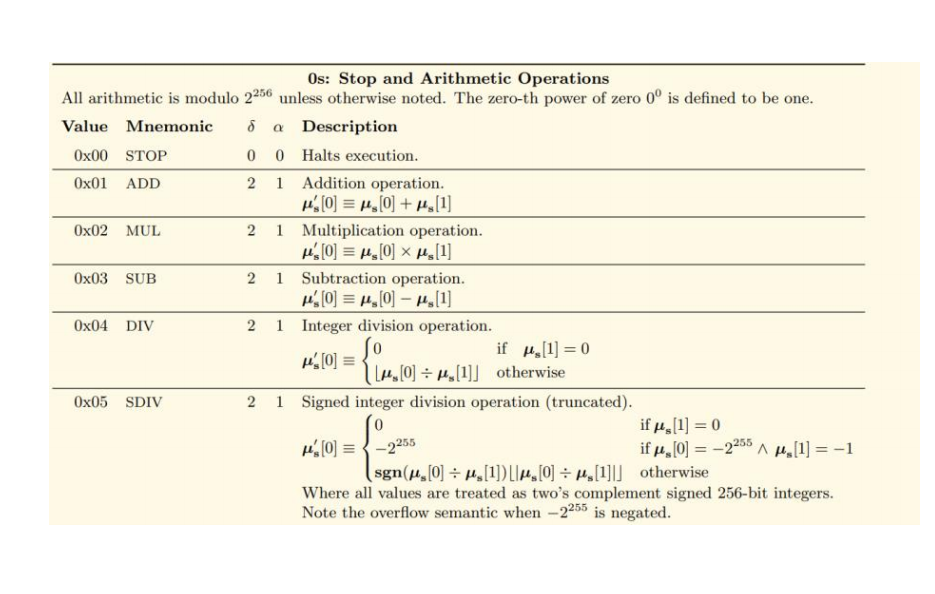}
\caption{Correspondance of bytecode and opcode.}
\end{figure}

\begin{figure} 
\centering
\includegraphics[scale=0.55]{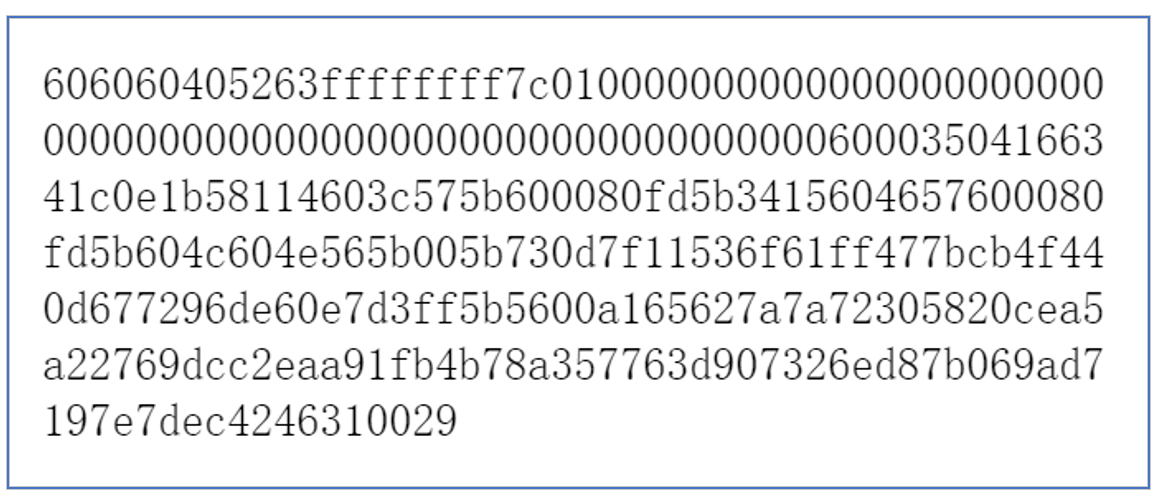}
\caption{Sample bytecode of an smart contract}
\end{figure}

\begin{figure}[!h!t]
\centering
\includegraphics[scale=0.58]{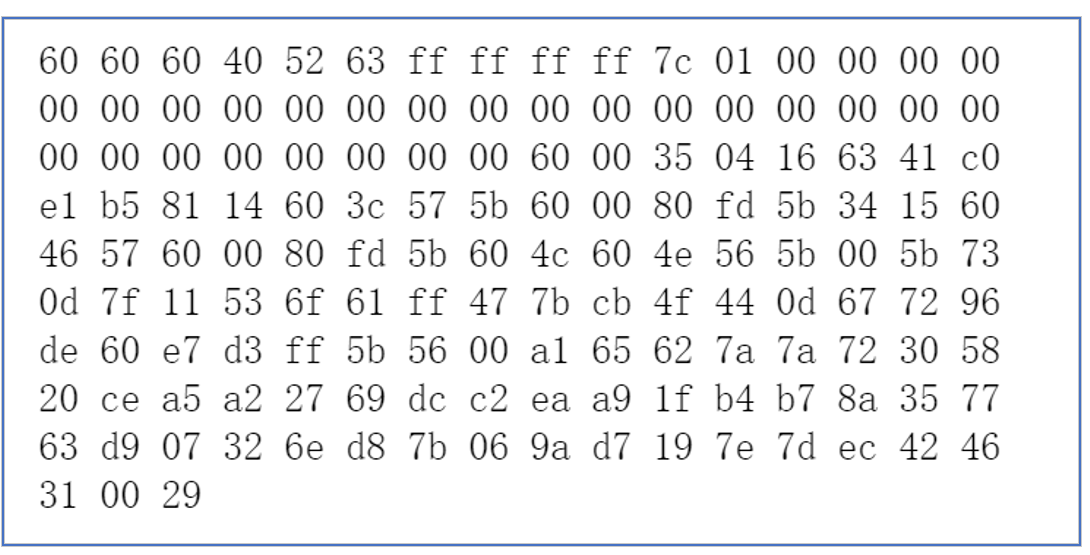}
\caption{Sample bytecode after splitting into two characters.}
\end{figure}
\subsection{Feature Description}

\begin{figure*}[ht]
\centering
\includegraphics[scale=0.45]{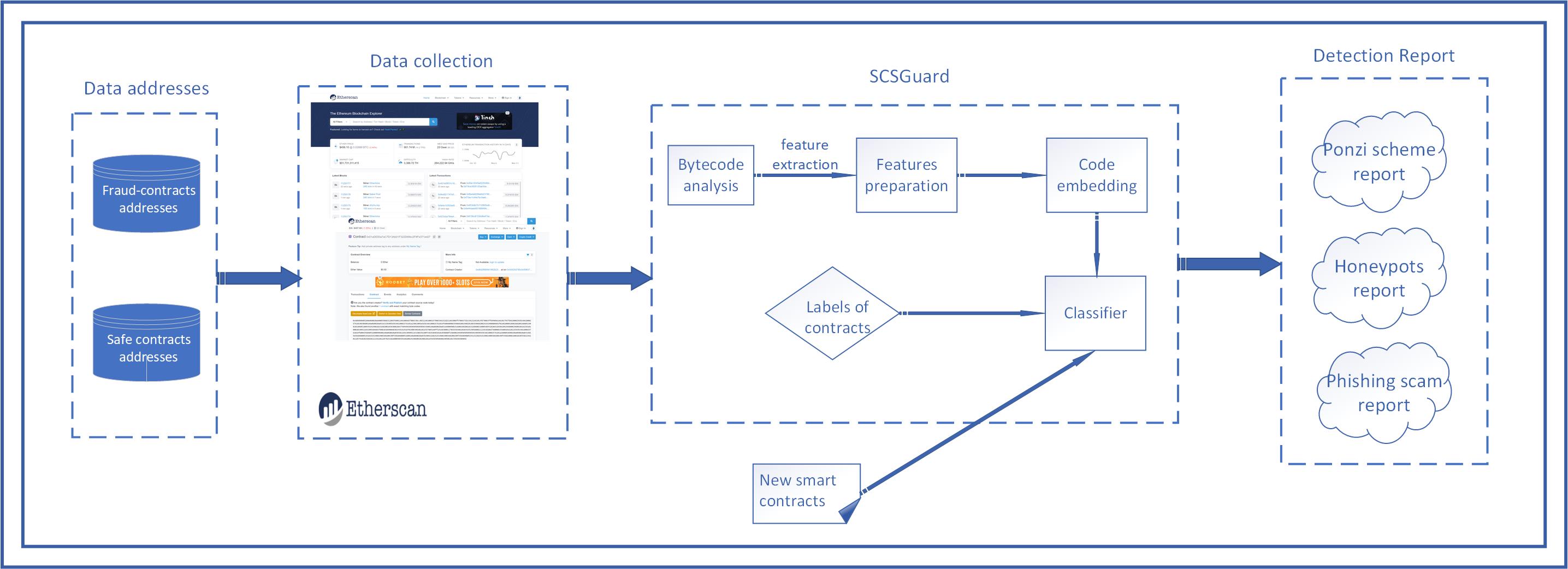}
\caption{The overview of our approach}
\end{figure*}

Fraud detection based on machine learning algorithms needs appropriate and informative features. However, initial feature selection for this task is particularly laborious and challenging due to the absence of strong natural and obvious features. Defining patterns of scams through domain knowledge is a desirable method, but it is hard to cover various types of new scams, let alone some malicious adversaries that may intentionally break standard rules. Therefore, in this work, we do not predefine any patterns for fraud-related smart contracts.

Ethereum smart contracts usually have transaction history and code open to the public. All transaction data will be recorded on Ethereum due to the nature of blockchain. However, transaction data is growing continuously, and any contract could send a transaction to another contract at any time. Hence, it is difficult to use transaction data to detect frauds in smart contracts. Code includes source code and bytecode, but source code is not available for all contracts. As a result, we focus on the analysis of EVM bytecode of smart contracts which can be obtained directly from the blockchain even if the corresponding source codes are unpublished.

To exploit treats of smart contracts bytecode, we need to convert them to a vectorial representation first. The bytecode of a smart contract, as shown in Figure 1, is a string of low-level instructions. Unfortunately, it is hard to find the inside features based on the raw bytecode, since the lengths of the bytecode of smart contracts can vary from several lines to thousands of lines and it is hard to figure out the inside features.

However, according to Ethereum yellow paper \cite{Wood}, each value corresponds to an opcode (operational code), as it is shown in Figure 2, there are 256 different opcodes defined to have specific meanings. To take advantage of this feature, we need to split the bytecodes into a series of bytes each contains two characters that each byte can reveal some features of smart contracts. Figure 3 and Figure 4 gives an example of the transformation of smart contract.

These bytes need to be converted into a vectorial representation before being used as inputs of our algorithms. Byte $n$-grams have been commonly used as features in many works in natural language processing and its application in malicious code detection is attractive in works \cite{Jurafsky}\cite{Jang}. Inspired by this, we use byte N-gram method to transform them into vectorial representation by regarding the unique combination of every $n$ consecutive bytes as an specific feature. The size of vocabularies (number of distinct $n$-grams) extracted for the bytecode n-grams representation is of 257, 19600, 67148, 150898, for 1-gram, 2-gram, 3-gram and 4-gram respectively. Later, the bytecode $n$-grams are transformed into vectors, which served as the inputs of the following machine learning classifiers.

\section{Implementation}
In this section, we first give a brief introduction of our approach and then describe the implementation details. In a high-level picture, SCSGuard takes the bytecode of a smart contract as the input. In terms of the output, SCSGuard generates a bug report of the input smart contract.

\subsection{Overview of SCSGuard}
Our approach aims to explore methods of machine learning techniques to create accurate detectors for scams in smart contracts. The methodology can be summarized into four phases: Data Collection, Sequence Preprocessing, Smart Contract Analysis and Detection Report. The structure of our algorithm can be seem in Figure 5.

For the first phase, we crawl the bytecode of smart contracts using the open public Website Etherscan, as has been discussed in section \Rmnum{3}. During the second phase, SMRdetector splits the bytecode into a two-character set. Then, the n-gram features of bytecode will be generated. After the two preparation phases, we use the features and labels of smart contracts to train a classification model. The last step is to generate the fraud detection report of a smart contract.

The overall process of designating smart contracts as malicious or benign is divided into the training, validating and testing phases. First of all, a training-set of smart contracts is provided to the classifier model. Each bytecode of the contract is parsed, and a vector representing each contract is generated. The vectors of the contracts and their labels are the input for the learning model. By processing these vectors, aiming to minimize the loss function, a classifier is trained. Next, during the validation process, the high-level parameters of the classifier will be modified. Then, in the testing phase, a collection of new contracts which did not appear in the training and validation phase will be classified by the learning model. Finally, the performance of the generated learning model is evaluated by metrics, including accuracy, precision, recall and F1-score.

\subsection{Feature Attention Mechanism}
We propose a classifier based on attention mechanisms to detect scams in smart contracts. The overall architecture of our Attention Network is shown in Figure 6. In this section, we first describe the use of the attention mechanism in our architecture.

\begin{figure}
\centering
\includegraphics[scale=0.45]{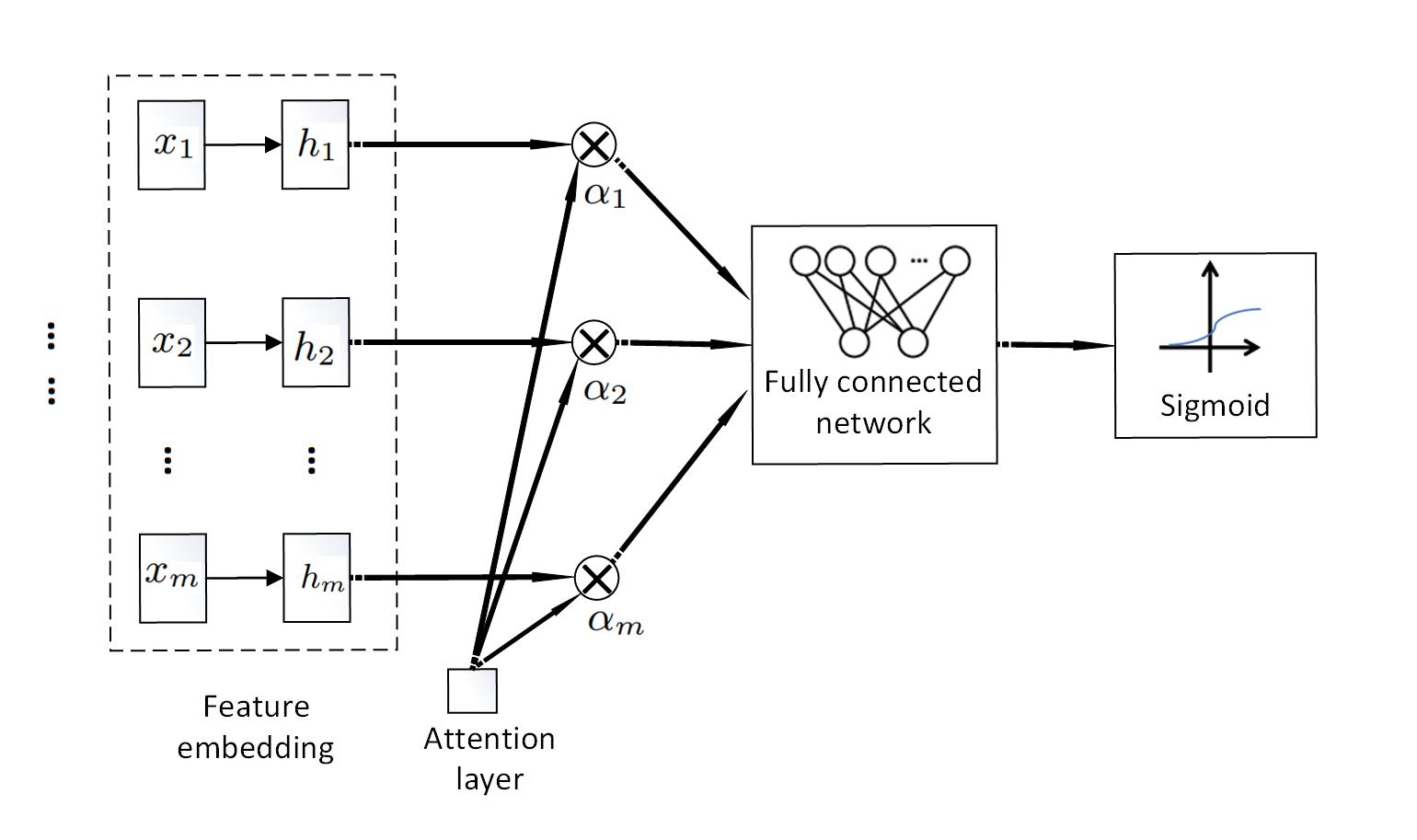}
\caption{Classification architecture with attention}
\end{figure}

After feature extraction in Section \Rmnum{3}, we get the $n$-gram features of smart contracts. However, not all features contribute equally to the representation of a smart contract. When applying $n$-gram features to neural networks, there is a high number of the features; thus, it will be challenging to learn the overall structure of smart contracts. Hence, during the detection process, we introduce an attention mechanism to extract such features that are important to the meaning of smart contracts.

There are various attention mechanisms in the field \cite{Luong}, in this work, we mainly focus the analysis on scaled dot-product attention \cite{Vaswani}, which is one of the most common use attention mechanisms. Consider an input token sequence of length $m: \boldsymbol{x} = x_1,x_2,... ,x_m$, where  $x_i$ is the $i$-th input token whose representation before the attention layer is $\boldsymbol{h}_{i} \in \mathbb{R}^{d}.$. The token representation $\boldsymbol{h}_{i}$ can be the output of an encoder or the word embedding. An intuitive understanding of attention function it to map a query and a set of key-value pairs to an output, where all the query, keys, values, and output are vectors [32]. 

Then, the $attention\ score$ for the $i$-th token which captures the $absolute$ importance of it is:
\begin{equation}
    a_{j}=\frac{\boldsymbol{h}_{i}^{\top} \boldsymbol{V}}{\sqrt{d}},
\end{equation}
where ${\sqrt{d}}$ is a commonly used scaling factor used to normalize the dot products, and $\boldsymbol{V} \in \mathbb{R}^{d}$ is the context vector that can be viewed as the value obtained from a fixed query with the input sentence (key). 

The corresponding $attention\ weight$ which indicates the $relative$ importance can be calculated as:
\begin{equation}
    \alpha_{i}=\frac{\exp \left(a_{i}\right)}{\sum_{i^{\prime}} \exp \left(a_{i^{\prime}}\right)}
\end{equation}
Then, the $i$-token after the attention layer can be represented as $\alpha_{j} \boldsymbol{h}_{i}$, which can be feed into the next layer.

During the training section, assume we have $N$ labeled smart contracts, then the training set can be represented as: $\mathcal{D}=\left\{\left(x^{(i)}, y^{(i)}\right)| i = 1,2,..., N\right\}$.Where $x^{(i)} \in \mathbb{R}^{d}$ is the extracted features associated with the $i$-th smart contract, $y^{(i)}\in\{0,1\}$  is the verified label of the contract. If the contract is a verified as a scam, the label $y$ will be 1. Otherwise, $y$ will be equal to 0.

Unlike traditional classification models that usually return the class label directly, the last layer of our model is a Sigmoid function, and the output will be a probability between 0 and 1. If the predicted probability is smaller than 0.5, the contract will be considered as a safe contract. Otherwise, it will be recognized as a scam.

The objective of our classifier is to determine whether a given contract is a scam. In the training process, we aim to minimize the following objective function:
\begin{equation}
Obj(\theta)=L(\theta)+\Omega(\theta),
\end{equation}
where $L$ is the training loss, and $\Omega$ is the regulation term. 

The training loss function $L(\theta)$ measures the sum of losses on the training data. Since the loss provided by each training data point, we expect the classifier model can learn from the errors. 

Loss function plays an essential role on the performance of the model, so it is usually selected specifically for its application. The most common ones are squared loss, softmax, and cross-entropy loss (logarithmic loss). In this work, considering our model is a binary classification, we choose the cross-entropy loss function:
\begin{equation}
L(\theta) = -\frac{1}{N} \sum_{t=1}^{N} \left[y^{(i)} \ln \left(\hat{y}^{(i)}\right)+\left(1-y^{(i)}\right) \ln \left(1-\hat{y}^{(i)}\right)\right]
\end{equation}

As for the regulation term $\Omega(\theta)$, which can help to prevent overfitting by penalizing the complexity of the model, we define it as:
\begin{equation}
\Omega(\theta)=\frac{1}{2} \lambda \|\boldsymbol{\omega}\|^{2}
\end{equation}
where $\lambda$ is the parameter of the model and 
\begin{equation}
 \|\boldsymbol{w}\|^{2} \equiv \boldsymbol{w}^{T} \boldsymbol{w}=w_{0}^{2}+w_{1}^{2}+\ldots+w_{N}^{2} 
\end{equation}

As the network improves its estimation accuracy, the cross-entropy loss tends towards zero, indicating that during the training process, the model gradually grows to be more accurate in classifying smart contracts, it minimizes the distance between the desired output $y$ and the output estimate $\hat{y}$.

\subsection{Performance Measures and Validation}
To evaluate the performance of classification models, the test accuracy, Eq (7), is the simplest metric which measures the fraction of predicted correctly class among all instance.
\begin{equation}
    Accuracy = \frac{true\ positive + true\ negative}{total\ tested}
\end{equation}

\begin{figure*}[t]
\centering
\includegraphics[scale=0.72]{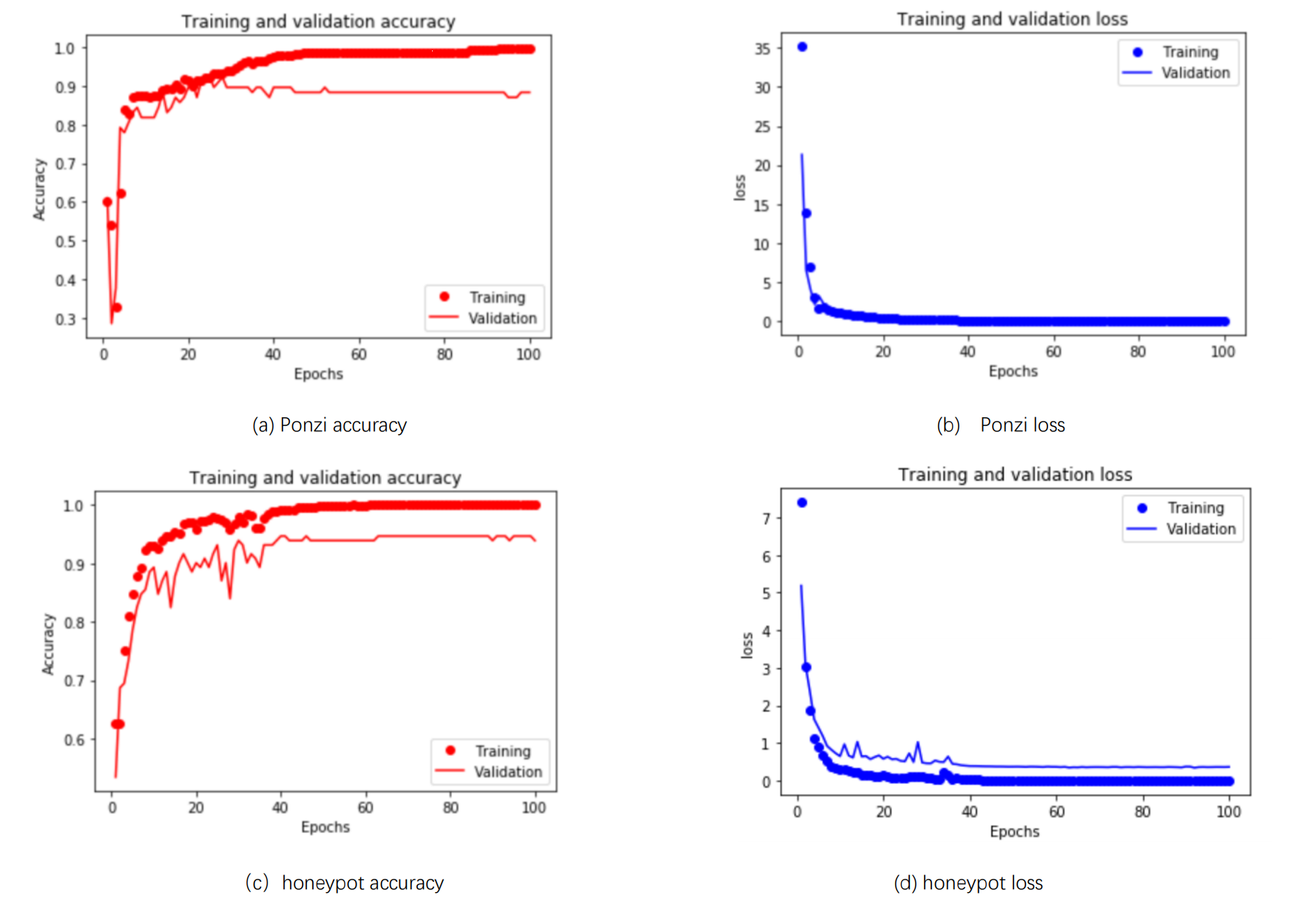}
\caption{The performance(accuracy and loss) of SCSGuard on Ponzi scheme and honeypot detection over 100 epochs.}
\end{figure*}

However, we face the problems with imbalanced class distributions, among all the verified contracts, the malicious ones only occupy a small part. We denote the majority class as the negative class and the rare class as the positive class. In the imbalanced distribution problem, the accuracy metric losses its effect, because if the classifier predicts all instance as a negative class, it can still achieve a high accuracy. Therefore, we further check the performance of classification models using three other metrics including precision score, recall score and F1-score. We denote a positive instance correctly classified as $true\ positive$, a positive instance wrongly classified as $false\ negative$; a negative instance correctly classified as a $true\ negative$ and a negative instance being classified wrongly as a $false\ positive$.

\begin{equation}
    Precision = \frac{true\ positive }{true\ positive + false \ positive}
\end{equation}
\begin{equation}
    Recall = \frac{true\ positive }{true\ positive + false \ negative}
\end{equation}
\begin{equation}
    F1-score = 2 \times \frac{Precision \times Recall}{Precision + Recall}
\end{equation}

The precision metric measures the ability of a model not to mislead safe smart contracts as a scam, and the recall metric can measure the ability of a model to find all negative instances. We also present F1 score in equation (10), which is the harmonic mean between precision and recall F1.

\section{Experiments and Results}
In this part, we illustrate the performance of SCSGuard using three kinds of scams. We also evaluate the performance of SCSGuard by involving different constructions of neural networks and compare it with other state-of-the-art classification models.

\subsection{Performance of SCSGuard}
Our proposed classification model-SCSGuard, has 689,473 parameters in all. The classifier result is based on a binary sigmoid activation function. In the training process, we found that our model was easier to train on a balanced dataset. We first divided the data into 60\% training, 20\% validation, and 20\% test. Then we undersample 1000 unique safe smart contracts randomly from the dataset. Also, we oversample the negative training smart contracts to an equal number to obtain a balanced training dataset. 

We evaluated the performance of our model on the task of detecting Ponzi schemes and honeypots respectively. The performance of SCSGuard is shown in Figure 7. 
\begin{table}[!htb]
    \caption {Ponzi scheme detection measures}
    \centering
    \begin{tabular}{|l|l|}
        \hline
        \thead{Classfication Performance Measure} & \thead{SCSGuard(\%) } \\
        \hline
        \makecell[c]{Test Accuracy} & \makecell[c]{0.922}\\
        \hline
        \makecell[c]{Precision Score} & \makecell[c]{0.963}\\
        \hline
        \makecell[c]{Recall Score} & \makecell[c]{0.978}\\
        \hline
        \makecell[c]{F1-Score} & \makecell[c]{0.971} \\
        \hline 
    \end{tabular}
    \label{bs}
\end{table}

\begin{table}[!htb]
    \caption {honeypots detection measures}
    \centering
    \begin{tabular}{|l|l|}
        \hline
        \thead{Classfication Performance Measure} & \thead{SCSGuard(\%) } \\
        \hline
        \makecell[c]{Test Accuracy} & \makecell[c]{0.947}\\
        \hline
        \makecell[c]{Precision Score} & \makecell[c]{0.942}\\
        \hline
        \makecell[c]{Recall Score} & \makecell[c]{0.989}\\
        \hline
        \makecell[c]{F1-Score} & \makecell[c]{0.964} \\
        \hline 
    \end{tabular}
    \label{bs}
\end{table}

During the test session, as shown in TABLE \Rmnum{1} and TABLE \Rmnum{2}, SCSGuard can achieve 92.2\% accuracy, 97.1\% F1-score in the detection of Ponzi schemes, and 94.7\% accuracy, 96.4\% F1-score in the detecion of honeypots. The evaluation measures show that SCSGuard can achieve high scores in fraud detection tasks.

We also test SCSGuard on a small set of phishing scams, the result shows that SCSGuard can find 4 in 5 phishing scams. It proves the potential of SCSGuard on finding unknown scams.

\subsection{Analysis of the Architecture of SCSGuard}
There are several critical designs in SCSGuard, including $n$-gram features in the contract embedding layer, and the attention layer which used to align the hidden states produced by the networks. In this part, we provide an intuitive understanding of the construction of SCSGuard. The experiment results show that the $n$-gram feature embedding and the attention mechanism are two essential parts of SCSGuard.

\subsubsection{Contribution of N-gram Feature Embedding}
To use bytecode information to detect scams in smart contracts, we need to convert bytecode into vectors before feeding into our learning models. We compare the results with different feature embedding. The first neural network model without $n$-gram feature selection is named Feature Directly Embedding Method. In this method, we convert the bytecode of a smart contract into one-hot vector representations.

\begin{table}[!htb]
    \caption{comparison of different feature embedding methods}
    \centering
    \begin{tabular}{l|lll}
        \hline
        \thead{Methods} & \thead{Precision} & \thead{Recall} & \thead{F1-Score} \\
        \hline
        \hline
        Feature Directly Embedding &0.682&0.719 & 0.699\\
        1-gram Feature Embedding & 0.818& 0.820& 0.818\\
        2-gram Feature Embedding &\textbf{0.963}& \textbf{0.978}& \textbf{0.971}\\
        3-gram Feature Embedding &0.958& 0.979& 0.968\\
        \hline       
    \end{tabular}
    \label{bs}
\end{table}

We can observe that $n$-gram feature embedding significantly improve the performance of the classification model. If we use Feature Directly Embedding layer, the test accuracy is 0.682. However, with the $n$-gram feature added, the performance of learning model is significantly improved. Since the numbers of features for 1-gram, 2-gram, 3-gram are 257, 19600, 67148 accordingly, the time for training will increase similarly. So we choose n-gram feature embedding as our method without affecting the performance of SCSGuard.

\subsubsection{Contribution of Attention Mechanism}
There are a large number of parameters after the embedding layer, and not all parameters are equally important to the model. Attention mechanism, which improves the performance of many natural language processing tasks, can make the classification model pay more attention to more relevant and useful features. We add the attention layer after feature embedding layer.

To evaluate the function of attention mechanism in SCSGuard, we compare different neural network structures:
\begin{table}[!htb]
    \caption{function of attention mechanism}
    \centering
    \begin{tabular}{l|lll}
        \hline
        \thead{Network Architecture} & \thead{Precision} & \thead{Recall} & \thead{F1-Score} \\
        \hline
        \hline
        RNN & 0.858& 0.942& 0.898\\
        RNN+Attention &0.943& 0.976& 0.959\\
        LSTM & 0.818& 0.820& 0.818\\
        LSTM+Attention & 0.952& 0.978& 0.964\\
        GRU & 0.893& 0.976& 0.932\\
        GRU+Attention&\textbf{0.963}& \textbf{0.978}& \textbf{0.971}\\
        \hline       
    \end{tabular}
    \label{bs}
\end{table}

The performance of SMSGuard is improved by adding an attention layer. As is shown in TABLE \Rmnum{4}, GRU with attention mechanism can achieve the best evaluation results, so we use this neural network structure in our final classification model. 

\subsection{Comparison of Different Classification Models}
To evaluate the performance of our proposed algorithm, we also compare the performance with other classifiers to figure out which one is the most suitable algorithm for this classification task. These machine learning models including support vector machine (SVM) \cite{Hearst}, Random Forests (RF) \cite{Breiman}, and XGBoost \cite{Quinlan}. 

We compare the performance of these algorithms on the detection of Ponzi schemes and honeypots respectively:

\begin{table}[!htb]
    \caption{ponzi scheme detection performance}
    \centering
    \begin{tabular}{l|llll}
        \hline
        \thead{Algorithm} & \thead{Accuracy}& \thead{Precision} & \thead{Recall} & \thead{F1-score} \\
        \hline
        \hline
        SVM &0.831&0.957&0.800 & 0.871\\
        RF & 0.922& 0.978& 0.900& 0.937\\
        XGBoost &0.909&0.957& 0.898& 0.926\\
        SCSGuard &\textbf{0.922}&\textbf{0.963}& \textbf{0.978}& \textbf{0.971}\\
        \hline       
    \end{tabular}
    \label{bs}
\end{table}

\begin{table}[!htb]
    \caption {honeypots detection performance}
    \centering
    \begin{tabular}{l|llll}
        \hline
        \thead{Algorithm} & \thead{Accuracy}& \thead{Precision} & \thead{Recall} & \thead{F1-score} \\
        \hline
        \hline
        SVM & 0.761& 0.958 & 0.775& 0.857\\
        RF &0.953& 0.979& 0.913&0.945\\
        XGBoost &0.946&0.958&0.933&0.945\\
        SCSGuard &\textbf{0.947}&\textbf{0.942}&\textbf{0.989}& \textbf{0.964}\\
        \hline 
    \end{tabular}
    \label{bs}
\end{table}

We can observe that the performance of SCSGuard is better than other algorithms obviously. In performance metrics of SCSGuard, we can see that the recall score, which is the fraction of scams that have been uncovered accurately, is exceptionally high compared with other algorithms. It means our proposed algorithm will the least likely to miss scams, which is the most important part of fraud detection task.

\section{Conclusion}
Financial scams based on blockchain and cryptocurrency have become an important research problem. We have proposed a new approach based on the bytecodes of smart contracts to detect scams. We proposed an $n$-gram pattern-based learning algorithm with attention mechanism applied. We have demonstrated encouraging results of our method in detecting several popular scams. Our method achieves high-performance measures and outperforms other state-of-the-art classification models in the detection of Ponzi schemes and honeypots.  


\end{document}